# ProRuka: A highly efficient HMI algorithm for controlling a novel prosthetic hand with 6-DOF using sonomyography

Vaheh Nazari[1*], Yong-Ping Zheng[1,2*]

[1]Department of Biomedical Engineering, The Hong Kong Polytechnic University, Hong Kong. 999077, China; vaheh.nazari@polyu.edu.hk (V.N.); yongping.zheng@polyu.edu.hk (Y.-P.Z.)

[2]Research Institute for Smart Ageing, the Hong Kong Polytechnic University, Hong Kong, 999077, China; yongping.zheng@polyu.edu.hk (Y.-P.Z.)

[*]Corresponding author: vaheh.nazari@.polyu.edu.hk and yongping.zheng@polyu.edu.hk

**ABSTRACT:**

*Sonomyography (SMG) is a novel human-machine interface that controls upper-limb prostheses by monitoring forearm muscle activity using ultrasonic imaging. SMG has been investigated for controlling upper-limb prostheses during the last two decades. The results show that this method, in combination with artificial intelligence, can classify different hand gestures with an accuracy of more than 90%, making it a great alternative control system compared to electromyography (EMG). However, up to now there are few reports of a system integrating SMG together with a prosthesis for testing on amputee subjects to demonstrate its capability in relation to daily activities. In this study, we developed ProRuka, a novel low-cost 6-degree-of-freedom prosthetic hand integrated with the control provided by a SMG system with a wearable ultrasound imaging probe. The classification of hand gestures using different machine learning classification/regression algorithms including KNN, nearest neighbor regression, random forest, decision tree classifier, decision tree regression, support vector regression and support vector machine in combination with a transfer learning model (VGG16) was first evaluated off-line to determine its reliability and precision. Additionally, the developed controlling system were evaluated on two amputees, in real-time experiments using a variety of hand function test kits. The results from an off-line study including ten healthy participants indicated that nine different hand motions can be classified with a success rate of 100%. In addition, the hand function test in real time (using 4 different hand gestures) confirmed that the designed prosthesis with the SMG controlling system can assist amputees to perform a variety of hand movements needed in daily activities.*

## 1. INTRODUCTION

Hands help to perform the majority of human activities of daily living, and losing one or both hands will result in independence reduction [1]. Even though most artificial limbs used today are either purely cosmetic or serve a practical purpose with limited functionalities (such as a hook-like gripper), various multi-fingered prosthetic hands have been developed and commercialized [2-4], including the i-Limb Hand, KIT hand, Michelangelo Hand, Bebionic Hand, and Vincent Hand, all of which depend on electrical motors and complex mechanical components. These neuroprosthetic hands have limited use for amputees due to their hefty weights (>400 g) and high cost (from $10,000 to $75,000) [5-8]. Moreover, the invention of additive technology revolutionized the manufacturing methods by decreasing the cost of production and the weight of the robot, as well as speeding up the product development process. This invention also affects the industry of prosthesis, encouraging researchers and engineers in creating numerous 3D printed prosthetic hands [9-14].

Despite the advancements in developing novel, dexterous, and state-of-art prosthetic hands with the ability to assist amputees in performing different daily activities [15-17], around 50–70% of patients refuse to wear and use the current prosthetic hand due to its poor functionality, high cost [5-8, 12, 18], low comfort, lack of sensory feedback and most importantly not accurate controlling system, not being able to effectively predict user's intended movements and provide natural-like control over prosthesis [15, 19].

To identify the most important features of upper-limb prosthesis, several studies have been conducted. The key factors can be listed as anthropomorphic characteristics (kinematics, size, weight, and appearance) [3, 20], performance (speed, force, and dexterity) [21, 22], and strong and integrative grasping [5, 23, 24]. Bioinspired motion speeds and an adequate grip force are necessary for the device to be useful for carrying out activities of daily living (ADL) [25]. However, among the most fundamental needs for a robotic prosthesis is the capability to control the robot with sufficient precision and responsiveness of the fingers [22], so that it may be used effectively and with sufficient dexterity [5, 26-28].

Till now, numerous projects have been conducted to develop novel and dexterous prosthetic hands, such as "Revolutionizing Prosthetics", "CyberHand", "SmartHand", "Bebionics", "Luke", "Neurobotics", and others, for improving the functionality of robotic hands [12]. However, most of these prosthetic hands still have various limitations, including bulkiness or not enough force provided, to be utilized in ADL [29].

Moreover, despite the study of various human-machine interfaces (HMI), there is still a lack of prosthetics with reliable control of multiple degrees of freedom [15]. For instance, using biological signals such as electromyography (EMG) or electroencephalography (EEG) as a non-invasive approach has been studied and proposed as a popular HMI, enabling users to control rehabilitation devices not only for prostheses but also rehabilitation robots [30] and exoskeletons [31, 32]. However, these techniques are very noisy and the recorded signals can be affected by electrode movements as well as sweating [33]. Also, EMG sensors are not able to monitor deep muscles activities, making this controlling system unable to be used in predicting more complex hand gestures with acceptable accuracy. For EEG control, the response time is still relatively slow [34-36]. In addition, the intended hand gestures perform by robot limited, and the most commercialized EMG-controlled prostheses still only have close and open functions, although different approaches for controlling robots with high dexterity have been proposed at the research level.

In recent years, in order to improve the quality of signals recorded from sensors as well as decrease the amount of noise, invasive techniques such as implanted EMG, targeted muscle reinnervation, myoelectric implantable recording arrays (MIRA) [37], magnetomicrometry (MM) [38], and others have been proposed. However, invasive approaches raise numerous questions regarding safety and efficacy since the electrodes need to be implanted into the body [33]. The field has been searching for a signal which can represent individual muscle activation and be collected noninvasively.

Over the last two decades, using signals extracted from the ultrasound images of muscle during contraction to control prosthetic hands has been a popular research topic. Zheng et al. first studied the feasibility of controlling robotic hands using an ultrasound device in 2006, in which the term "sonomyography" (SMG) was proposed by the team for this non-invasive HMI approach [39]. Basically, SMG refers to the signal representing architectural changes of a muscle detected by real-time ultrasound images during its contraction [40]. Since ultrasound imaging can inherently differentiate the activities of both deep and superficial muscles as well a group of neighboring muscles simultaneously and non-invasively, SMG has attracted the attention of a lot of researchers since it was proposed [41-45]. Recently, a unique mobile SMG system to monitor muscles' activities was evaluated regarding its reliability and validity by Ma et al. in 2019, paving the way for real-time monitoring of muscle activity throughout both indoor and outdoor activities especially for controlling prostheses using a wireless SMG system [46].

A number of SMG-based prosthesis control systems have been earlier reported in the literature, which mainly focused on the demonstration of feasibility, including using single element transducers [47 49]. A low-power SMG system designed for wearable use with a prosthetic hand was proposed by Engdahl et al. in 2020 [48]. Using AI for classifying different intended hand gestures, authors demonstrated that their suggested technique successfully classified nine distinct finger motions with an accuracy of around 95%. In 2020 Yang et al. [49] advocated the use of wearable 1D SMG in combination with subclass discriminant analysis (SDA) and principal component analysis (PCA) to predict wrist rotation (pronation/supination) and finger movements. This research demonstrated that the SDA machine learning method could be used to identify both finger gesture and wrist rotation concurrently with an accuracy of around 99.89% and 95.2%, respectively.

To overcome difficulties caused by single element transducer, a number of studies reported the use of B-mode imaging transducers [47]. In a study published in 2019, Akhlaghi et al. [50] evaluated the effect of using a sparse set of ultrasound scanlines to determine the optimal location on the forearm for capturing the maximal deformation of the primary forearm muscles during finger motions and classifying various types of hand gestures and finger movements. The results indicated that the ultrasonic probe should be placed over around 40–50% of the forearm's length in order to identify distinct hand motions with greater precision. This is because the largest muscle activation occurs at this region. In addition, the categorization result demonstrated that employing B-mode ultrasound to operate a prosthetic hand is a viable option, since the accuracy was almost 95%. In 2019, Li et al. [51] tested the capabilities of M-mode and B-mode ultrasound to detect 13 various hand and finger movements in 8 able-bodied subjects. Using the SVM algorithm to classify various hand gestures, the accuracy of the M-mode classification was determined to be 98.83±1.03%, and the B-mode classification was determined to be 98.77±1.02%. On the other hand, the accuracy of the Backpropagation Artificial Neural Network (BP-ANN) classifier was 98.77% in M-mode and 98.76% in B-mode. They discovered that M-mode SMG transducers were equally as accurate as B-mode SMG signals when it came to detecting wrist and finger movements as well as differentiating between a variety of hand gestures, which suggests their possible utility in human-machine interfaces.

Zheng et al in 2006 [39] and Guo et al. in 2008 [52], for the first time conducted experiments to evaluate the relationship between morphological changes of forearm muscles and the wrist angle. The results of their study showed that muscle deformation measured by ultrasound is correlated linearly with the wrist angle. Moreover, in 2011 and 2012, Castellini et al [53, 54], conducted an exciting experiment to assess the potential of SMG system in predicting the position of the fingers using ultrasound images collected from human forearm. The result of their study, by discovering a linear relationship between finger position and extracted features from ultrasound images, showed that this novel controlling system has a great potential for not only predicting the intended hand gestures but also providing information regarding the finger position and amount of flexion, enabling SMG controlling system to provide proportional and natural like control experience to people with amputation.

For a more complete understanding of the various systems and techniques developed using ultrasound imaging of muscle or SMG for controlling upper limb prostheses, readers can refer to a review paper recently published by Vaheh et al. (2023), which conducted a comprehensive evaluation and comparison of the results and findings of previously published works on the SMG system as a novel human-machine interface. The outcomes of this review paper demonstrated the promise of ultrasonic sensing as a practical human-machine interface for the control of bionic hands with multiple degrees of freedom. In addition, this review showed that a variety of machine learning algorithms combined with feature extraction models could correctly classify various hand gestures with an accuracy of about 95% [55, 56].

Despite all the feasibilities demonstrated about using SMG together with machine learning or deep learning methods for detecting hand gestures with the potential for prothesis control, there are few reports about testing the SMG controlled based robot on real amputees [57]. Considering that residual muscles after amputation surgery are very different from those in normal subjects, the promising results demonstrated in earlier papers may not necessarily stand for residual limbs. In addition, up to now there is still no report of a system integrating SMG together with a prothesis for testing on amputee subjects to demonstrate its performance in daily activities.

In this study, we report the novel SMG controlling system, design and performance of a lightweight (360 g), functional and cost-effective 6DOF prosthetic hand called ProRuka (Figure 1). The novel ProRuka was developed and tested by considering anthropomorphism, functionality, safety, and comfort, all of which were inspired by the structure of the human hand. To evaluate the accuracy of the proposed machine learning model to classify different hand gestures needed in daily activities, ten able-bodied volunteers were recruited to attend our first experiment. For the off-line evaluation experiment, the data were first collected from the ten able-bodied participants before being used to train the model and assess the accuracy of AI model. Among all the data collected from the ten volunteers, around 70% of them were used for training and the rest for validation. For the amputee subjects, the data were collected from the individual residual limb and used for training their individual models. This step is very similar to the training session for using conventional EMG controlled prothesis. The trained model together with the prosthesis and the controlling system was evaluated with two amputee subjects, who performed standardized hand function tests including the Box and Blocks (B&B) test, Targeted Box and Blocks (TB&B) test and Action Research Arm Test (ARAT).

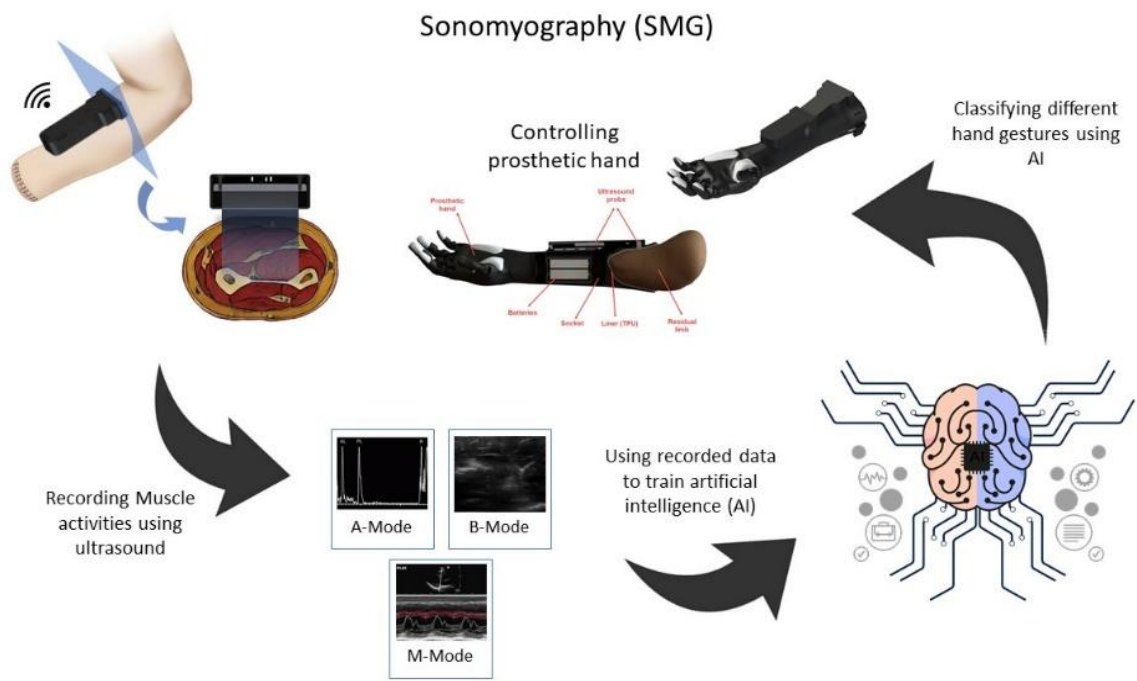

**Figure 1 Sonomyography as a novel HMI method:** The overall schematic of the sonomyography (SMG) and controlling prosthetic hand using an ultrasound probe.

## 2. SYSTEM AND AI MODEL DEVELOPMENT
### 2.1. Classification of different hand gestures using ultrasound imaging

For the control part reported in this paper, different classification methods were studied. Participants were divided into able-bodied and amputee groups. Each volunteer was asked to sit in a comfortable position and put their hand on a cushion. Then, the muscle activities in different hand gestures were captured using a palm-sized wireless ultrasound probe. Before classification, CNNs were used to extract features from each image, and these features were used to train a model with a machine learning algorithm including RF, KNN, DTC and SVM or regression methods including decision tree regression (DTR), nearest neighbor regression (NNR), and support vector regression (SVR) with 3 different kernels, Linear (SVR-L), and Polynomial (SVR-P). Then, the accuracy of each machine learning algorithm was examined and compared.

### 2.1.1. Feature Extraction

To classify the finger movements and gestures more actively and effectively, since machine learning algorithms cannot process all raw information contained in the images, a CNN algorithm with pre-trained weights in this study was used to extract features from each images before using collected data to train the AI model. Three different pre-trained models including VGG16, VGG19 and InceptionResNetV2 were individually used for feature extraction. To select and extract features, 64 filters from the first convolutional layer were utilized. The features extracted from the training data were then used for classification. It is important to note that more filters using be used to extract more features, but training the model would require more time and GPU memory.

### 2.1.2. Classification

Figure 2 shows the overall schematic of the whole classification process. After extracting features, these data were used for training three different machine learning classification algorithms including RF, KNN, DTC and SVM as well as four regression methods (DTR, NNR, SVR-L, SVR-P) to classify different hand gestures and finger movements. Two-thirds of the collected data were used for training and the rest were applied for validation.

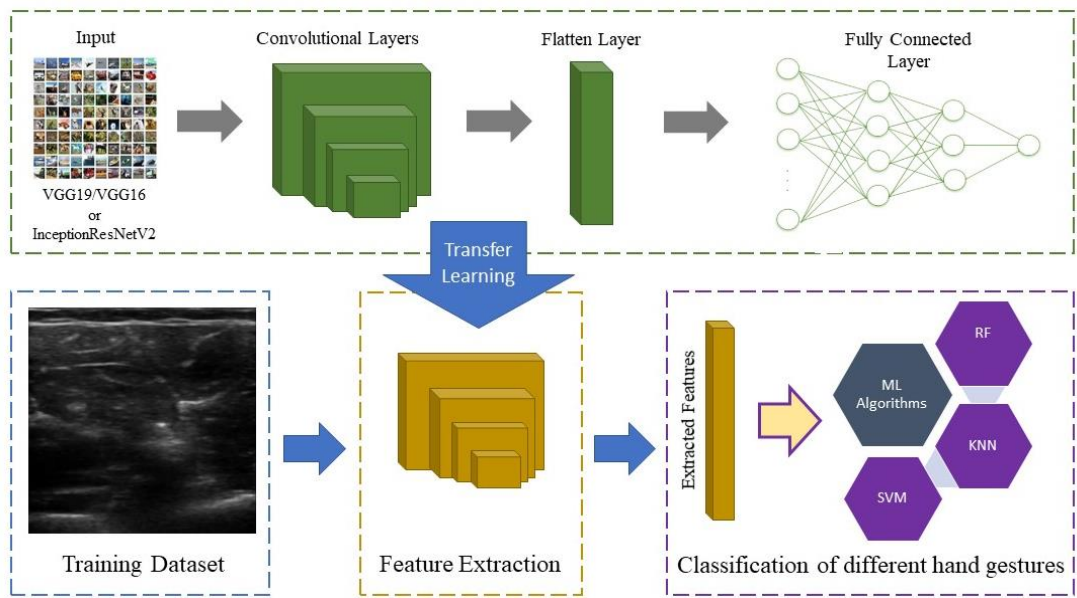

*Figure 2 The overall schematic of the classification process:* *A transfer learning model was used to extract features from images and the extracted features were then utilized for training the model using a machine learning algorithm.*

### **2.2. Replacing ultrasound gel and gel-pad with a sticky silicone pad**

For the sticky silicone pad, biocompatible silicone liquid (Deping, Guangdong Province, China) was used to create a pad using the molding technique. In the experiment, two different silicones with hardness ratings of Shore 00-00 and Shore 00-05 were utilized. Three different silicone pads were created for the experiment. The first one was a silicone pad with a hardness of 0. The ultrasound images had a good resolution using this pad, but it was too sticky, and it was difficult to put it on the hand with the prosthesis. A second silicone pad with a hardness of 05 was created. The resolution was good for controlling the prosthesis, but the pad was fragile and could be damaged easily during donning and doffing. Thus, for the third pad, a combination of silicones with 00 and 05 hardness were mixed together in a 3:1 ratio. The testing results demonstrated that the image quality provided by this silicone gel pad was good enough for controlling the robot, and it was sticky enough to minimize transducer movement. Additionally, the flexibility of the pad was good enough to be used with a socket without any damage (Figure 3).

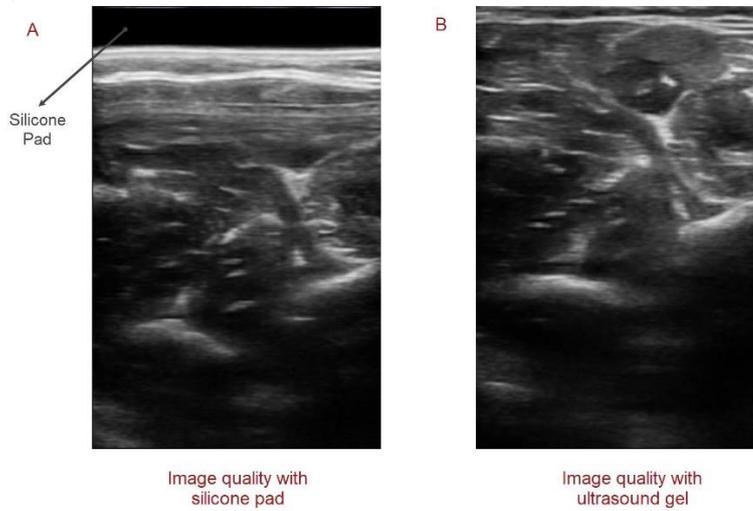

**Figure 3 Utilizing a silicone pad instead of ultrasound gel:** Image quality using a silicone pad (A) ultrasound gel (B)

## 2.3. Designing a novel prosthetic hand

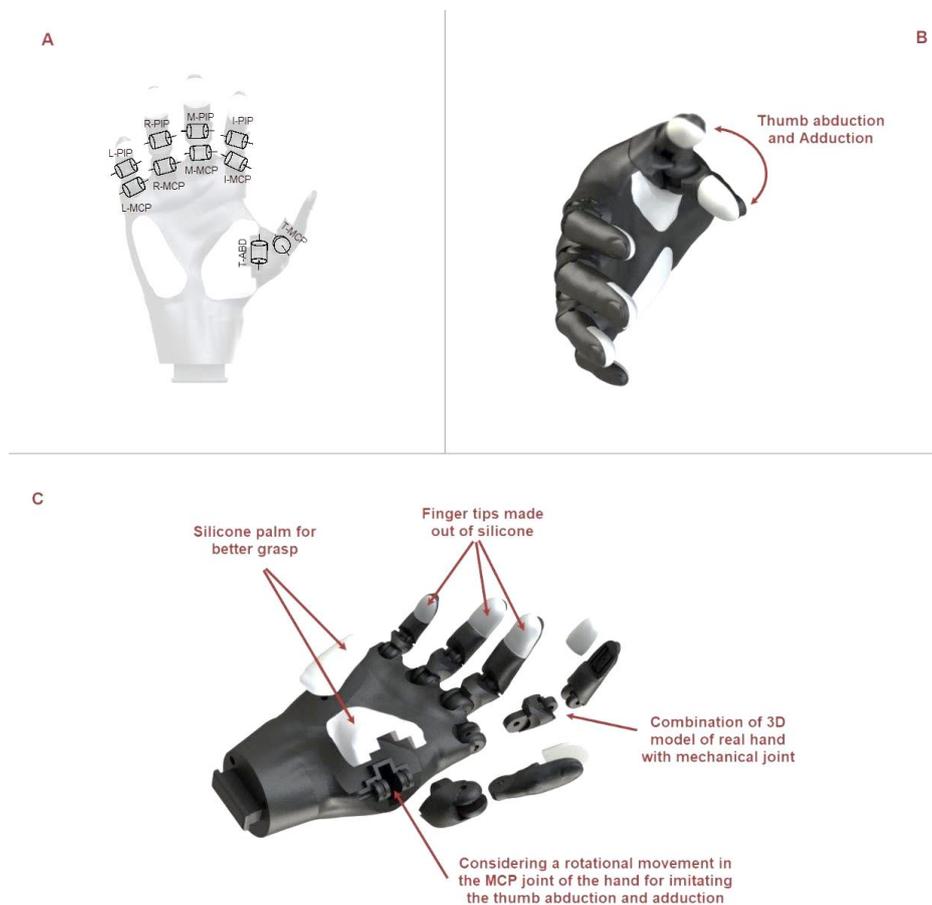

**Figure 4: ProRuka a 6DOF prosthetic hand:** A) The front view of the prosthetic hand. B) Additional rotational joint in the CMC joint for mimicking thumb abduction and adduction. C) Exploded view of the prosthesis

The objective of designing the prosthesis in this study was to develop a low-cost, lightweight, user-friendly, dexterous, multiple degree of freedom, functional prosthetic hand to help amputees perform functional and needed hand and grasping gestures for ADL. In order to make the prosthetic hand resemble a normal human hand, a 3D model of a normal human hand was first prepared using a portable industrial 3D scanner (EinScan Pro 2x, Shining 3D, Hangzhou, Zhejiang, China). Then mechanical joints were replaced with the hand joints of the scanned model to make the prosthesis functional (Figure 4). It is vital to mention that thumb abduction and adduction play an important role in grasping different types of objects and perform 80% of daily living hand activities. Consequently, we considered rotational movement in the MCP joint in order to make a prosthesis that can perform thumb abduction and adduction (Figure 4B). In order to increase the friction between objects and the prosthesis and decrease the chance of objects slipping and falling from the prosthesis, the fingertip of each finger as well as the palm of prosthesis was made of silicone. Furthermore, 3D printing technology was utilized to make the prosthesis cost-effective and lightweight.

### 2.3.1. Silicone

Each finger and palm were made of two different materials: black nylon and silicone. The plastic item was printed using an additive manufacturing method (VPrint 3D, Hong Kong, China). However, for the silicone part, a mold for each item was designed and printed using a 3D printer with black nylon material and the final form of the different parts of the prosthesis was created by a molding process using a Platinum Cure Silicone Rubber Compound with a shore hardness of 00-50 (Smooth-On, Macungie, PA, USA).

### 2.3.2. Actuation system

Inspired by human hand anatomy, an artificial tendon mechanism was used to flex and extend each finger [5, 58]. To flex each finger and perform thumb adduction movement, fishing wires (Sufix, Greensboro, NC, USA) were attached to the fingertip on one side and the roller of the DC geared motor with a stall torque of 200 g.cm and rotational speed of 185 rpm (Fuzhou Bringsmart Intelligent Tech. Co., Ltd, Fuzhou, China) on the other side, passing through the button side of the finger. Each finger could be flexed by the rotation of the motor shaft by pulling the artificial tendon. A fishing wire on the other side was connected to a tension spring (RS PRO, London, UK), allowing for finger extension and thumb abduction. Once the finger is flexed, the tension spring stores the energy and releases it when the motor is driven in the other direction, causing the finger to extend.

To make the whole prosthetic hand compact, a motion-control unit was designed to contain all actuators and microprocessors inside it which was fitted into the prosthesis. Under each motor's shaft, a groove on the unit was drilled to lead the movement of the fishing wire and convert the rotational movement into prismatic movement (Figure 5).

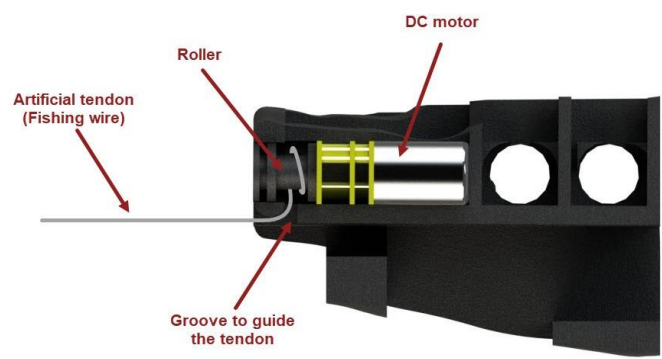

**Figure 5: The mechanism for finger flexion:** A groove on the motion-control unit for leading the fishing wires to the roller connected to the shaft of the motor as well as converting the rotational movement of the motor's shaft into linear movement.

### 2.3.3. Control unit

The prosthetic hand, in addition to the actuators, contained the Arduino Nano (Arduino LLC, Italy), a Bluetooth module (HiLetgo, Shenzhen, Guangdong, China), five sensors (Honeywell, Charlotte, NC, US), and three dual op-amps (Texas Instruments Inc., Dallas, Texas 75243 USA) (Figure 6), while the socket contained the ultrasound probe, two 3.7V 2200mAh batteries, six DRV8871 motor drivers, and two Wowoone voltage boosters.

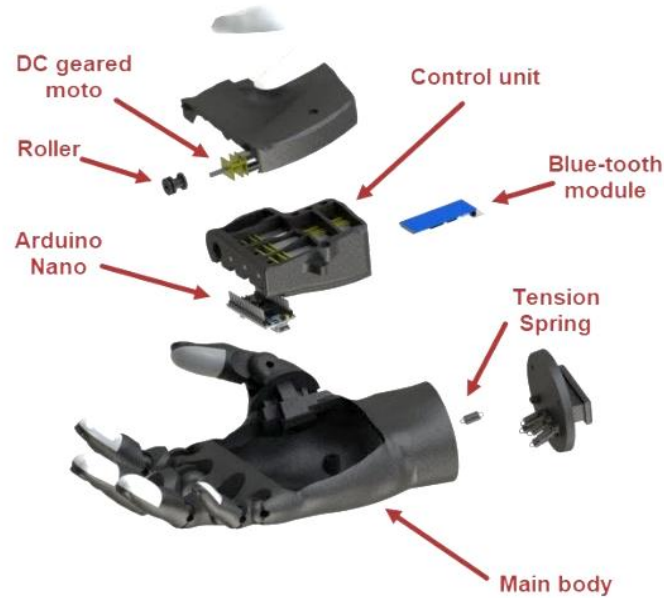

**Figure 6: Exploded view of the prosthesis:** Different electrical components mounted into the prosthetic hand

To control the robot, captured images using an ultrasound probe were first sent to the computer through Wi-Fi. The received images were then processed by a machine learning model, and the predicted result (the prediction of a hand gesture based on each image) was sent to the Arduino via Bluetooth. The microprocessor then, based on the predicted result, could perform different hand gestures by sending a trigger to the related motor driver. A sensor was placed on top of each finger to measure and control the fingertip force. To measure the amount of force, before connecting the sensor to the Arduino, the output voltage of the force sensor was amplified by an LM358 op-amp. In total, three dual op-amps were used for five sensors. To provide the power to run the motors, two 3.7-volt rechargeable batteries were utilized. To increase the voltage of each battery, a voltage booster was utilized to provide 12 volts for each motor driver. Nevertheless, to provide the power to turn the Arduino on, a voltage booster was used to provide 5V from the battery of the ultrasound probe, which used 3.7V.

### 2.3.4. Force sensor

In order to control the amount of force provided by the robot, a force sensor was mounted on top of each finger. A silicone coating was applied to the tip of each finger to increase the friction between it and objects. Underneath the silicone, the force sensor was mounted. To transfer the applied force to the silicone, a curved surface model printed with black nylon material was fixed into the silicone (Figure 7A). It was necessary to design a mold for each fingertip so that round-shaped plastic could fit into the silicone (Figure 7B).

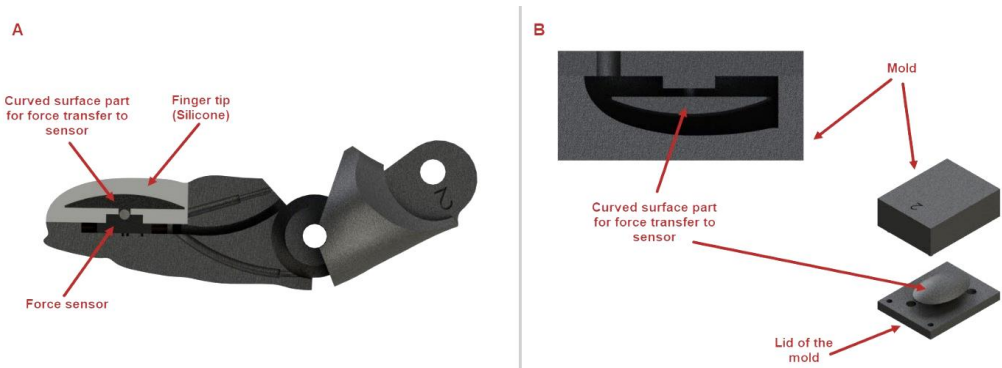

**Figure 7: The location of the force sensor in the fingertip:** A) The force sensor is mounted in the fingertip of each finger and a plate with a curved surface is fixed in the fingertip to transfer force to the sensor. B) A mold designed to fix the plate inside the fingertip made of silicone.

Since the output voltage of the sensor was not high enough to be measured by Arduino, an inverting amplifier was used to amplify the output voltage of the sensor. A LM358 (Dual Operational Amplifier) was utilized to amplify the output voltages of two sensors and in total three LM358 ICs were used in the prosthetic hand. The Arduino's 5-volt output served as the input voltage for each sensor, while each operational amplifier was powered by a battery (with the voltage increased to 12 volts via a booster). The output voltage after amplification was measured by the Arduino to convert the output of the sensor to the force applied to the fingertip (Figure 8).

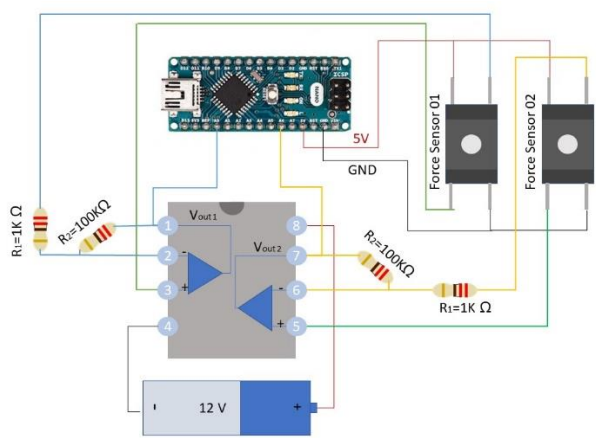

**Figure 1: Schematics of fingertip force measurement:** Sensors are powered by 5 volts of the Arduino's output voltage, while the LM358 is powered by a battery with a booster. The output voltage of each sensor was measured by Arduino after being amplified by an operational amplifier.

### 2.3.5. Socket design and fit

The stump acts as the major point of contact between the prosthetic device and the residual limb, which is called the socket. Suction is created between the residual limb and the prosthesis by the use of the socket, which is wrapped around the end point of the residual limb. To design a socket to fix the prosthesis to the residual limb of an amputee, the amputated hand of volunteers was scanned using a 3D scanner. Mainly, the socket was divided into two parts with two different materials. The first part was the main body, printed using black nylon material on a 3D printer, inside which the batteries, ultrasound transducer, and electronic components were mounted, and the second part was a linear part printed using soft TPU material with the shore A hardness of 50, placing it between the socket and hand to decrease the stress on the hand and to make the socket comfortable for the amputee (Figure 9).

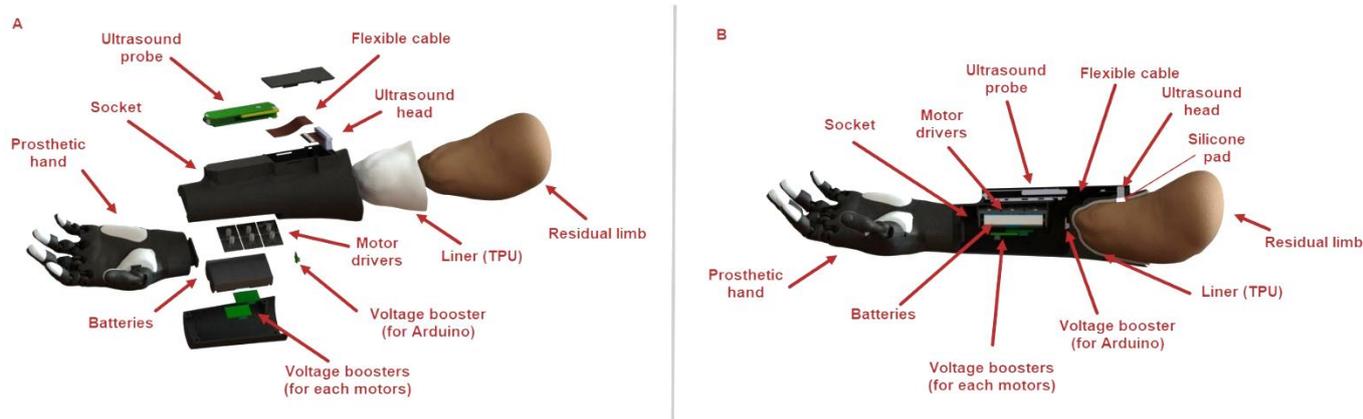

**Figure 9: The design of the socket:** A) Exploded view of the socket, showing the different electrical components inside the socket. B) A cut view of the socket showing the location of electrical components inside the socket.

## 3. EXPERIMENT AND RESULTS

### 3.1. Participants

Since the musculoskeletal anatomy is different in able-bodied people and those with transradial limb loss, it was important to assess the accuracy of the proposed classification method for both groups. Consequently, we separated the participants into able-bodied and amputee groups. The study was approved by the Human Subjects Ethics Sub-committee of The Hong Kong Polytechnic University (HSEARS20220720001).

Ten able-bodied volunteers (five males and five females, aged between 22 and 33) were recruited in this study (the healthy group). All participants had no hand impairments and disabilities. Two amputees (2 males aged between 45 and 69, referred to as A1 and A2) were recruited in this study (the amputee group). Both amputees were left-hand trans-radial amputees 26 and 45 years after their injury, respectively. Each participant completed an informed consent form after receiving information about the research and the experimental design.

### 3.2. Experimental Setup

The volunteers were asked individually to sit in a comfortable position and put their hand on a cushion and keep their palm upwards. A B-mode lightweight (only 67 g) wireless ultrasound module (Model UL-1C, Beijing SonopTek Limited, Beijing, China) was fixed on the forearm using a customized case. To collect maximum muscle activities the probe was placed perpendicular (transverse) to the forearm on 30% to 50% of the length of the forearm from the elbow (Figure 10). Moreover, to minimize the effect of transducer relocation on accuracy, data were collected at different transducer locations.

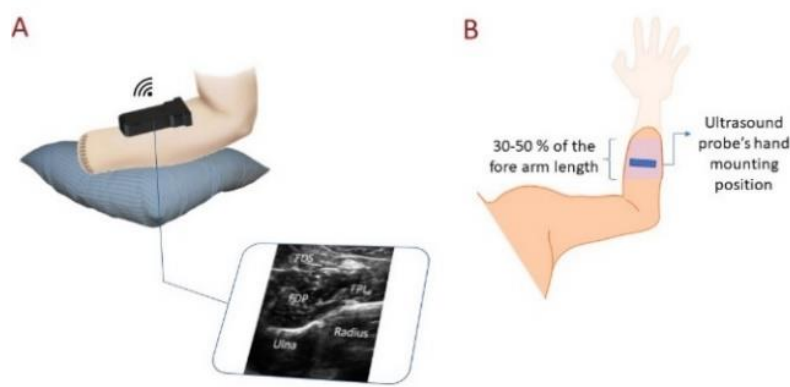

*Figure 10: Experimental setup: A) The experimental setup and the ultrasound image of the main muscles responsible for finger flexion. B) The area on the forearm to capture the best muscle activities to control the robot.*

### 3.2.1. Evaluating the developed prosthetic hand

In the first stage, in order to evaluate the potential of SMG as a novel HMI method, both off-line classification experiments were conducted in both able-bodied and amputee groups. The accuracy of the classification method with different machine learning algorithms including KNN, SVM and RF were compared. In the second stage of this study, physical properties and range of motion were measured and assessed. Moreover, fingertip and palmar grasp forces were measured, and the results were compared to those of other commercially available prostheses. Furthermore, the reliability of the sensors mounted on the fingertip was evaluated. In the final stage, the functionality of the developed machine learning model as well as prosthesis was evaluated by two volunteers with transracial hand amputation using different hand function evaluation kits.

*3.2.1.1 Physical properties and range of motion measurement*

Anthropomorphic-associated factors, including size, weight, appearance, and range of motion (ROM), are crucial when developing a prosthetic device that best mimics the human hand's behavior and characteristics. The ROM of ProRuka was measured by SolidWorks (Dassault System, MA, USA) CAD software, and the results were compared with those of the human hand and other developed prostheses. Moreover, the size and weight of the prosthesis were measured, and the size of the developed prosthesis was compared with the 50th percentile human hand [59].

*3.2.1.2. Fingertip and palmar grasp force measurement*

Enough fingertip force to manipulate different objects without dropping them is crucial. To evaluate each fingertip force, the prosthetic hand was fixed to the experimental benchtop, and by running each individual motor to flex each finger, the maximum force produced by each finger was measured using a load cell (Hunan Tech Electronic Co. Ltd., Changsha, China). However, the palmar grasp force of ProRuka was measured by a digital hand dynamometer (Camry, CA, USA). Each fingertip force and palmar grasp force were measured four times for validation of the results. The results of the forces produced by the prosthesis were compared with those of other commercially available prostheses.

*3.2.1.3. Fingertip force sensor evaluation*

To evaluate the reliability of the sensor placed on the fingertip of each finger, we tested the mechanism in a custom benchtop setup (Figure 11). Each finger of the prosthesis was fixed to a holder to maintain the applied force on the load cell (Hunan Tech Electronic co. Ltd., Changsha, China). The force produced by the prosthesis measured by the sensor and the value was compared with the force measured by the fingertip sensor. In order to validate the reliability of the fingertip sensors, this experiment was repeated three times under six different fingertip forces. Moreover, R-squared value as well as mean squared error (MSE) was measured to assess the reliability of each sensor.

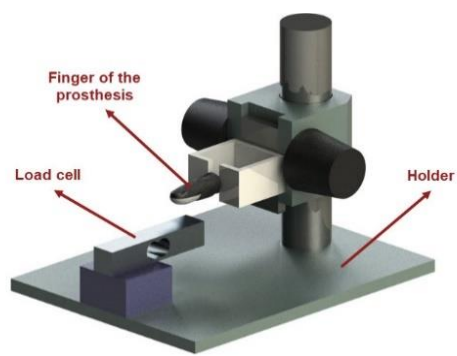

**Figure 2: Experimental setup:** Experimental setup for assessing the reliability of the force sensor mounted in the fingertip of the prosthesis

### 3.2.2. Experiment 1: Performance of off-line classification

Before conducting an experiment to evaluate the functionality of developed controlling system, an Off-line classification experiment was conducted in able-bodied group to evaluate the potential of SMG as a novel HMI method. The accuracy of the classification method with different machine learning classification/regression algorithms including DTC, NNR, DTR, KNN, SVR-L, SVR-P, SVM and RF were compared, after training and validation data collected from 10 able-bodied people. In the final stage, for further evaluation of the developed model, nested and non-nested cross validation were utilized.

*3.2.2.1. Data collection for off-line Test*

In the off-line test, intact groups were asked to sit comfortably on a chair and place their elbow on a pillow, with the palm facing upward. Before collecting data for training and validation, the position of the ultrasound transducer was first defined and fixed, making sure that key muscles, including flexor digitorum superficialis (FDS), flexor pollicis longus (FPL), and flexor digitorum profundus (FDP), are covered by transducer. Each subject then was asked to perform one of nine different hand gestures, including rest, individual finger flexion (index, middle, ring, little and thumb), fist, pinch and key pinch, and hold it for 5 seconds. All the 9 hand gestures repeated 3 times. To avoid fatigue and spasm in the muscles, there were 15 seconds of rest between each hand gesture. In the off-line testing of the able-bodied group, in total 11,625 images were collected and 8,350 of them were used for training, while 3,275 images (384*400 pixels) were used for validation.

### 3.2.3. Experiment 2: Real-time functional performance

To evaluate the functionality and performance of the reliability and dexterity of the proposed controlling system, different hand function tests were conducted. In this study B&B test as well as was TB&B test, which is a modified version of B&B test and action research arm test ARAT were utilized to evaluate the functional performance of prosthesis in daily activities. Before the evaluation session, two participants with transradial amputation were asked to attend two training sessions to improve their skills in controlling the robot as well as become familiar with the prosthetic hand and the process of the evaluation session.

*Box and blocks test:* Gross manual dexterity is often evaluated using a test called the B&B [60]. The evaluation kit consists of a box with two squared compartments which are separated by a partition (Figure 12). One of the compartments was filled with 150 wooden cubes (25 mm3), combining in such a way that the blocks may be found to rest in a wide variety of positions. The number of blocks that were moved over the barrier in the allotted time of 60 seconds was how the test was scored. The participants were free to carry the blocks in whatever order they wanted, provided that their fingers passed the partition between the two compartments before releasing the block into the desired location.

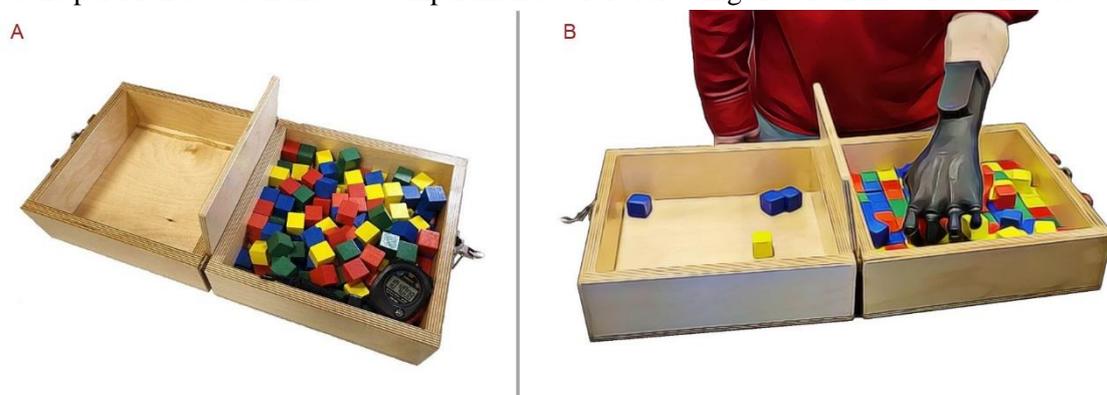

**Figure 12: The Box and Blocks test kit:** Using the B&B test kit to evaluate the hand function of the developed prosthetic hand

*Targeted box and blocks test:* The TB&B test was performed with 16 (for 4x4 TB&B Test [61]) and 9 (for 3x3 TB&B Test [62]) blocks. The TB&B Test is an upper-limb functional task designed to elicit ecologically meaningful activities such as movement initiation, grasping, transporting, and controlled releasing of items. In addition to its use in assessing patients' functional improvement after undergoing rehabilitation, this test may also be used as an outcome measure in clinical studies of upper-limb transradial prosthetic devices [63]. A standard grid was placed on both sides of the compartment, and volunteers were asked to move each block to the other side of the compartment into its mirrored location. The box was turned upside down so that the outside area could be used for the assessment, which would make it simpler to complete and would also avoid the prosthetic hand from colliding with the box's walls (Figure 13).

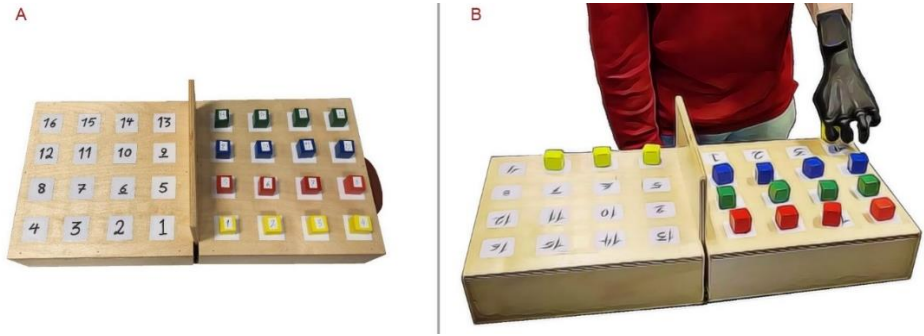

**Figure 3: The Targeted Box & Blocks test (4*4):** The modified version of the B&B test for assessing hand function

*Action research arm test:* The ARAT, which is extensively used to measure arm function, is one of the most prominent hand function evaluation kits. The testing kit consists of 19 different items to assess the different grasping types and arm movement (Figure 14). The whole assessment process takes approximately 10 minutes and scores are given based on the participants' arm movement and functionality and for each item the score is rated between 0 (no movement) and 3 (normal movement) [64, 65]. ARAT scores vary from 0-57, with 57 indicating higher performance. The final score can indicate weak (less than 10), moderate (10-56), and excellent (57) hand function [66].

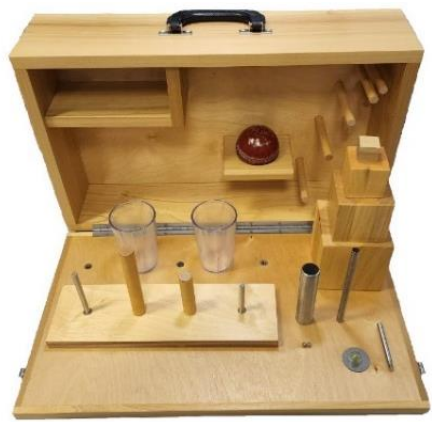

**Figure 4: The Action Research Arm Test kit:** Using the ARAT kit to evaluate the functionality of the developed 6DOF prosthesis and reliability of the machine learning model

### 3.2.3.1. Data collection for real-time classification test

In the real-time classification, to evaluate the whole SMG controlling system in the last session, different functional hand gestures including rest, pinch, key pinch, and cylindrical grip (fist) were classified as performing useful grasping types to help them use the robot in their ADLs. It is worth mentioning that out of four available grips, AI model was only trained with cylindrical grip, since the robot was not able to perform other grips.

During the experiment of real-time classification, the data were collected in two different dynamic and one static strategies, as the ultrasound image for each hand gesture would vary due to hand movements while performing different tasks. In static strategy, same as previous experiment, ultrasound images from forearm muscle were collected, while participants hand was placed on the table with the palm toward upward. In the first dynamic strategy, the participants were first asked to extend their hands and keep their palm in a supination position, then flip their hand without trying to move their wrist while performing and holding one of the hand gestures (Figure 15). This process was repeated three times for each hand gesture (rest, pinch, key grip, and fist). In the second dynamic strategy, volunteers were asked to extend their elbow and then rotate their forearm three times while performing and holding one of four hand gestures. Amputee subjects were asked to repeat this process twice. The whole process for each hand gesture took 120 seconds, a total of 480 seconds for four hand gestures.

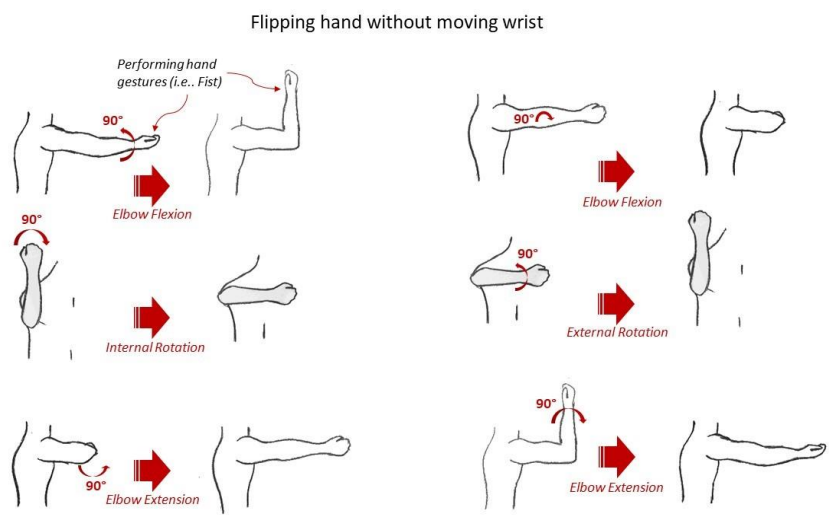

**Figure 15: Flipping hand without moving the wrist:** Flexing the elbow 90°, followed by 90° internal rotation and 90° elbow extension, then repeating the same movement with external rotation to flip the back hand

### 3.3. Results
### 3.3.1. Anthropometry of ProRuka

The results of the ROM, kinematics, and size of the prosthesis indicated high anthropometry. Figure 16A illustrates the direct comparison of the prosthesis size with the 50th percentile of the human hand [59]. The thumb's intermediate-distal diameter shows the most deviation from the reference hand model at 8.3%, demonstrating a high resemblance between ProRuka and the reference hand model. Regarding ROM and kinematics, Figure 16B shows that fingers 2–5 (index, middle, ring, and little) were able to flex 90° and 85° in the metacarpophalangeal (MCP) and proximal interphalangeal (PIP) joints, respectively. However, to have enough space to mount a sensor on the tip of each finger, the distal interphalangeal (DIP) joint was fixed. To perform different grasping types needed in daily activities, ProRuka was designed to be able to perform thumb abduction and adduction for better grasping. The developed prosthetic thumb can be positioned in abduction or adduction along a 64° range of motion. Figure 16B depicts the results of the comparison of the ROM and kinematics of the developed hand with the human hand and three different commercialized prostheses, including iLimb, Bebionic, and Michelangelo [3, 67]. The weight of the developed hand was 360 g, which is lighter than other commercialized prostheses (420–588 g) and the average human hand (400 g) [2, 3, 5, 17]. However, the socket that contains the batteries and ultrasound transducer weighs 420 g.

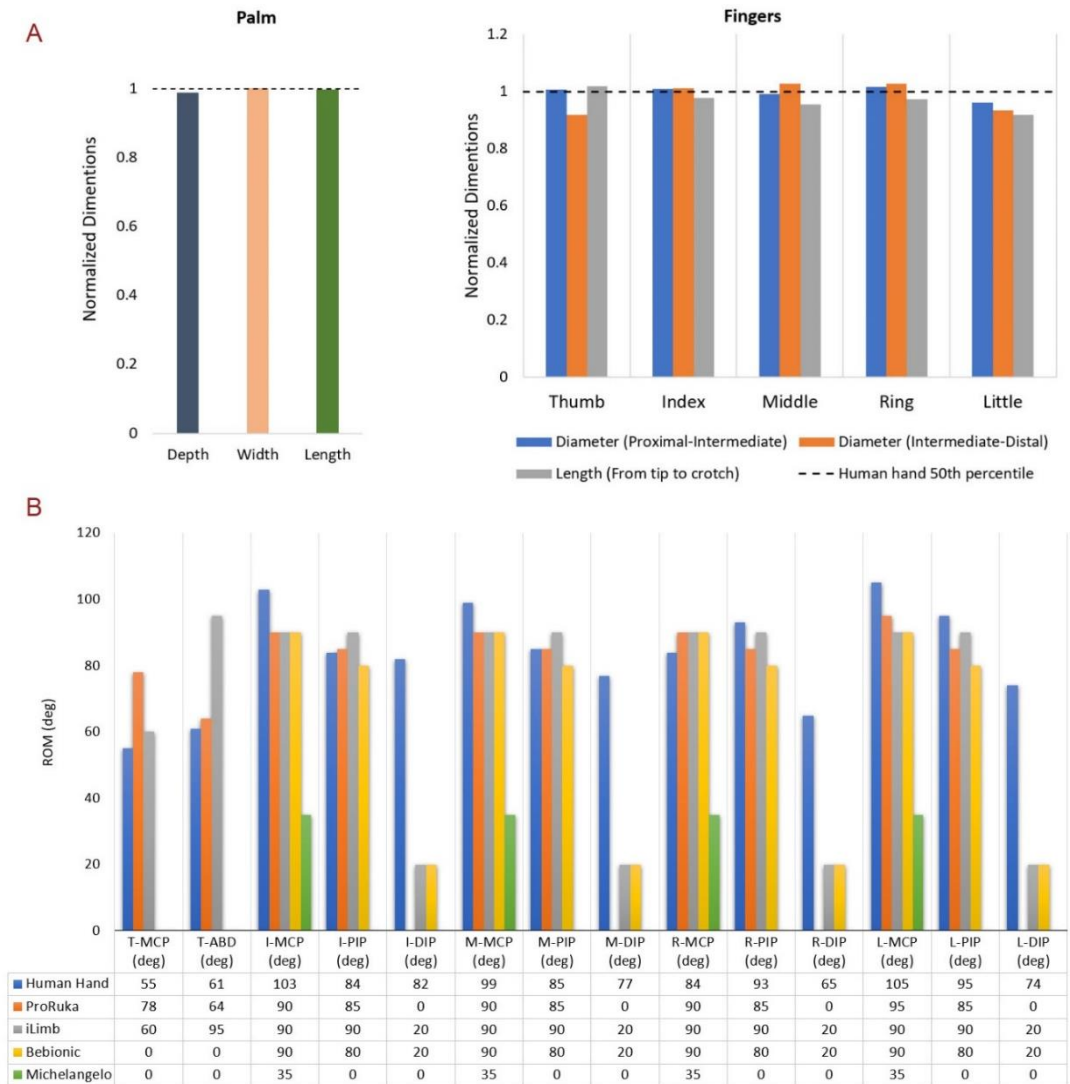

**Figure 5: Anthropometry of ProRuka:** Direct comparisons between ProRuka's size, ROM, and kinematic model and anthropometric data and the kinematics of a typical human hand and three commercially available prostheses including iLimb, Bebionic, and Michelangelo. A) Comparison of the size of ProRuka with a 50$^{th}$ percentile human hand. B) Comparison of kinematics and ROM of ProRuka with a human hand and three commercialized prostheses including iLimb, Bebionic and Michelangelo.

### 3.3.2. Fingertip, precision grip and palmar grasp force

The results indicate that the prosthesis is able to produce 2.7–5 N fingertip force and 17–18 N palmar grasping force. However, the amount of force can be increased by using motors with higher torque. The generated force is sufficient to hold a bottle weighing up to 780 g without tipping it over. Figure 17A illustrates the comparison of fingertip force of ProRuka for fingers 2–5 with four different commercial prostheses, including iLimb, iLimb Plus, Bebionic, and Vincent [3]. The comparison of palmar grasping force of the developed prosthesis with iLimb Plus, Bebionic, Bebionic V2, and Michelangelo is depicted in Figure 17B [3]. Figure 17C, however, shows the fingertip force of five fingers, precision grip force and palmar grasping force with the maximum standard deviation of 0.24 N.

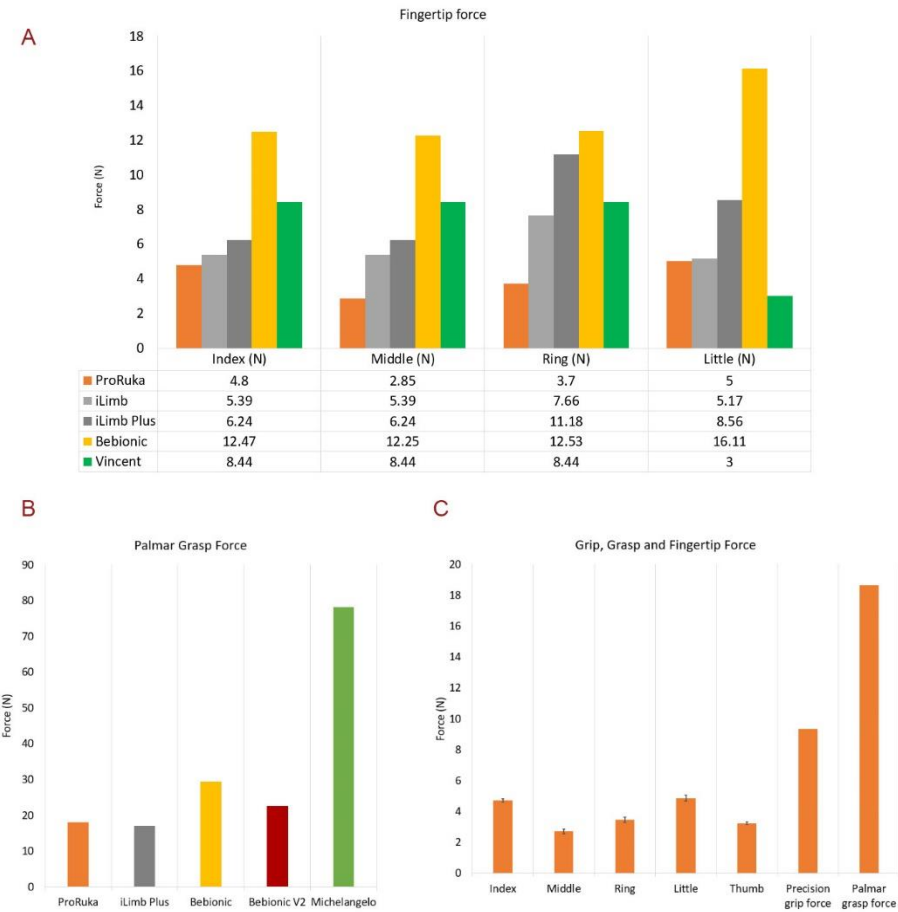

**Figure 17: Force experiment:** A) Comparison of the individual fingertip force of ProRuka with four different commercialized prostheses including iLimb, iLimb Plus, Bebionic and Vincent. B) Comparison of palmar grasping force of ProRuka with four different commercialized prostheses including iLimb Plus, Bebionic, Bebionic V2 and Michelangelo. C) Fingertip force of five fingers, precision grip force and palmar grasping force of ProRuka with the maximum standard deviation of 0.24 N.

### 3.3.3. Accuracy of fingertip sensors

To assess the reliability of the sensors the value of applied force was measured by sensors mounted on top of each fingertip and a load cell. Figure 18 illustrates the amount of force measured by sensors placed on top of each fingertip of the prosthesis compared with the actual value measured by the load cell. The R-squared value and MSE was 0.9844±0.0106 and 1399±458 g respectively.

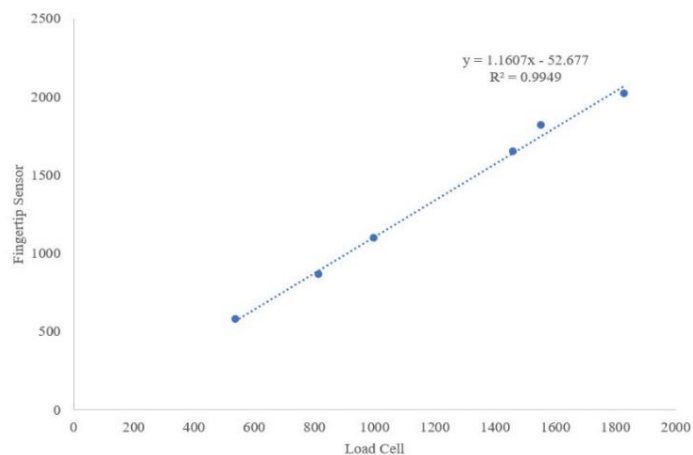

**Figure 18: The graph of measured force:** The amount of force measured by the sensor mounted on top of the finger (Fingertip Sensor) and Load Cell

### 3.3.4. Off-line classification results

Off-line classification results showed that the combining a transfer learning model with one of the machine learning classification algorithms (KNN, RF, SVM, DTC) and regression algorithm (NNR) are able to classify nine different hand gestures with an accuracy of 100% (Figure 19). Nevertheless, it was observed that more time was needed to train the model using SVR-L, SVR-P, DTC, NNR, DTR, and SVM while the RF and KNN were the fastest in training the model using collected datasets (around 205 seconds for the ten able bodied volunteer, with 8,350 images for training and 3,275 images for validation).

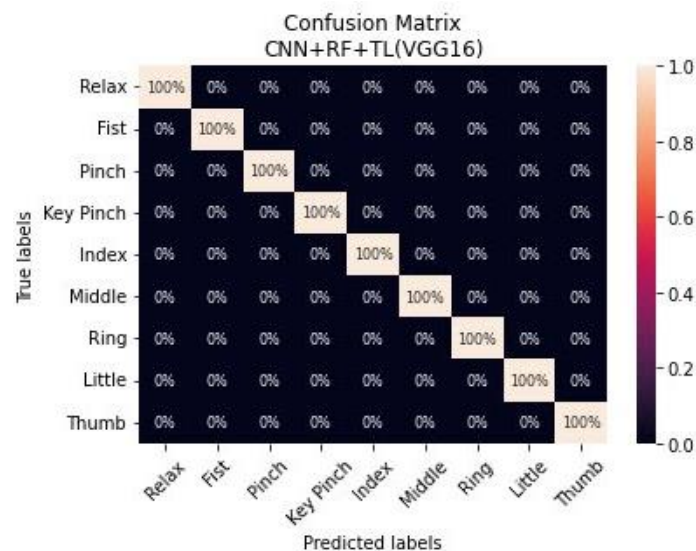

**Figure 19 The result of off-line test:** Off-line test results of classifying nine different hand gestures using different machine learning algorithms in the A) able-bodied group and B) the amputee group

### 3.3.5. Real-time performance results

Based on the off-line test results, VGG16 was used to extract features and a RF machine learning algorithm was utilized to train the model (the accuracy of classifying different hand gestures using this method was the highest). Two volunteers were invited to attend the experiment conducted to evaluate the functionality of the developed prosthesis. They were asked to complete the different hand function tests with the prosthesis in addition to their healthy hand to compare the results.

The final scores and results of the hand function test is summarized in Table 1. During the experiment, it was observed that a minimum of 120 seconds was needed to collect the training data for each hand gesture with an accuracy of 100%. It was also observed that the accuracy of classification was minimally reduced after transducer replacement due to donning and doffing the prosthesis. However, during data collection for training, data were collected at different transducer locations to minimize the effect of transducer relocation on accuracy.

The results of the B&B and TB&B tests showed that the volunteers were able to pick up the blocks by pinching, and during hand movements, no misclassification was observed. Both participants were able to easily transfer around 13 blocks without any training. However, during the TB&B test, the participants complained about the rigidity of the prosthesis.

The results of the ARAT showed that the developed hand had good performance in grasping and gripping different objects with different sizes but was unable to pick up small objects by performing a pinch gesture. Based on the scores earned by volunteers, the performance of the prosthetic hand was as good as a hand with moderate function.

### 3.3.6. Evaluating the potential of using a silicone pad instead of ultrasound gel or a gel-pad

In the experiment conducted to evaluate the potential of the silicone pad to be replaced with ultrasound gel to control the prosthesis using ultrasound imaging, we observed that a silicone pad can provide real-time images of the muscle with good image quality and that the captured data can be utilized to have real-time control over the prosthetic hand. Moreover, it was also discovered that the sticky silicon pad did not only stop the transducer relocation but also reduced stress on the skin by dampening the transducer's reaction force.

**Table 1: The result of hand function test:** The result of hand function evaluation using B&B, TB&B (4x4), TB&B (3x3) and ARAT tests

| Test | Hand | Missing hand | Result | |
|---|---|---|---|---|
| *B&B* | | | Number of blocks | |
| | | | A1 | A2 |
| | Left | Left | 12 | 8 |
| | Right | | 45 | 47 |
| *TB&B (4*4)* | | | Time (seconds) | |
| | | | A1 | A2 |
| | Left | Left | 86.66 | 136.79 |
| | Right | | 31.31 | 21.23 |
| *TB&B (3*3)* | | | Time (seconds) | |
| | | | A1 | A2 |
| | Left | Left | 41.40 | 67.18 |
| | Right | | 17.00 | 12.28 |
| *ARAT* | | | Score (Total) | |
| | | | A1 | A2 |
| | Left | Left | 40 | 40 |
| | Right | | 57 | 57 |

B&B: box and blocks, TB&B: targeted box and blocks test, ARAT: action research hand test.

## 4. DISCUSSION

Inspired by human hand anatomy a lightweight, functional and cost-effective prosthesis hand with 6DOF controlled by SMG called ProRuka was successfully developed. The anthropometry analysis of ProRuka showed that the functionality and size of the prosthesis are close to those of the human hand. However, in order to place force sensors in the fingertips of the prosthesis the DIP joint was designed to have a fixed angle. Nonetheless, it is worth mentioning that compared to other finger joints, the DIP joint has a small impact on the functioning of the hand [5]. Compared to other joints, the DIP joint has around four times less range of motion, as shown by Santina et al. That is why we kept this joint fixed in our design.

A complete comparison of the angular excursions of ProRuka's DOFs with the human hand and three anthropomorphic prostheses including iLimb, Bebionic, and Michelangelo prostheses showed that although the DIP joints were fixed, ProRuka's ROMs were quite similar to those of a human hand. Furthermore, compared to Bebionic and Michelangelo, ProRuka is able to perform thumb abduction and adduction enabling amputees to perform 80% of daily activities. Having a wider range of motion in the fingers, the prosthesis developed in this study, compared to the Michelangelo prosthesis, has better functionality.

Current commercialized prostheses typically are relatively heavy (420–588 g) and expensive (from $10,000 to $75,000). Hence to develop a lightweight and cost-effective prosthesis, 3D printing technology was utilized. By weighing 360 g, ProRuka is lighter than other commercialized prostheses and the average human hand (400 g) [2, 3, 5, 17]. On the other hand, the socket that contains the batteries and the ultrasonic transducer is 420 g in weight. Moreover, it is worth mentioning that the total production and material cost for ProRuka was USD828 excluding ultrasound probe, which was about USD1,000 for its module cost.

Kargov et al. conducted an experiment to analyze the fingertip force and grip force of the human hand to fulfill functional tasks. In their study they found that a maximum of 3.8 N of fingertip force and 16 N of grasping force was needed to hold an object weighing 522 g [68]. The results demonstrated that ProRuka is able to produce, on average, 3.79 N of fingertip force and, by producing a maximum of 9.38 N and 19 N of precision grip force and palmar grasp force, respectively, this prosthetic hand is able to hold and lift an object with a maximum weight of 780 g. Compared to other commercialized prostheses (iLimb, iLimb Plus, Bebionic and Vincent), ProRuka produces less fingertip force; however, by replacing actuators with motors with higher torque, the fingertip force of ProRuka will be increased if it is necessary. In Table 2, the information of commercialized prostheses in comparison with ProRuka is summarized.

**Table 2: Comparing ProRuka with commercialized prostheses:** Comparison of different functional prosthetic hands with the developed robotic hand [3, 5]

| *Prosthesis* | **Weight** *(g)* | **DOF** | *Number of Joints* | *Number of Actuators* | **Actuation system** | *Controlling System* |
|---|---|---|---|---|---|---|
| ***ProRuka*** | 360 | 6 | 10 | 6 | DC Motor: tendon-driven mechanism | SMG |
| *Hannes* | 480 | 10 | 14 | 1 | DC Motor: cable-based mechanism | EMG |
| *Michelangelo* | 420 | 2 | 6 | 2 | DC Motor: Worm Gear | EMG |
| *iLimb* | 460 | 6 | 11 | 5 | DC Motor: Worm Gear | EMG |
| *iLimb Plus* | 523 | 6 | 11 | 5 | DC Motor: Worm Gear | EMG |
| *Bebionic* | 588 | 6 | 11 | 5 | DC Motor: Lead Screw | EMG |
| *Bebionic V2* | 539 | 6 | 11 | 5 | DC Motor: Lead Screw | EMG |
| *Vincent Hand* | 509 | 6 | 11 | 6 | NA | EMG |

Despite the advancements in developing numerous prostheses, still developing a prosthesis with a reliable controlling system is challenging. However, in the recent years, SMG as a novel HMI method showed a great potential in control a prosthetic hand with a high accuracy by capturing the residual muscles' activities using ultrasound imaging. Figure 1 illustrates the SMG method as a new HMI technique for controlling prostheses with multiple degrees of freedom. In this study, the potential to control a prosthesis hand using SMG was evaluated. To classify different hand gestures, a combination of transfer learning models (including VGG16, VGG19 and InceptionResNetV2) and machine learning algorithms were utilized. And the results showed that this new method has high potential to be utilized in the control of prosthetic hands.

To train the model, different machine learning classification/regression algorithms, alone and in combination with a transfer learning model, were utilized. However, only a combination of the transfer learning model with one of the machine learning algorithms, including RF, KNN, DTC and SVM, as well as regression methods (NNR, and DTR) had the potential to classify nine different hand gestures with an accuracy of more than 91%. Table 3 presents a summary of the accuracy of different machine learning classification/regression algorithms in classifying 9 different hand gestures in off-line test. Moreover, Nested, and non-nested cross-validation (CV) scheme was applied to the offline experiment. The nested and non-nested CV score for offline test was 99.80% and 99.83% respectively. Furthermore, the accuracy of developed model in classifying different hand gestures of able-bodies and amputees, using RF and KNN algorithms was analyzed using ANOVA and Kruskal-Wallis test (KWT), and the results revealed no significant difference in classification accuracy among the groups ($p_1 = 0.36$ for ANOVA and $p_2 = 0.56$ for KWT), suggesting that the model performed similarly for both healthy individuals and amputees.

**Table 3:** *Accuracy of classifying hand gestures*: *The off-line classification of 9 different hand gestures in able-bodied group*

| Transfer learning | Machine learning algorithm | Accuracy |
|---|---|---|
| InceptionResNetV2 | RF | 100 % |
|  | KNN | 100 % |
|  | DTC | 100 % |
| VGG 19 | RF | 100 % |
|  | KNN | 100 % |
|  | DTC | 100 % |
| VGG 16 | RF | 100 % |
|  | KNN | 100 % |
|  | DTC | 100 % |
|  | NNR | 100 % |
|  | DTR | 91.72 % |
|  | SVR-L | 55.96 % |
|  | SVR-P | 55.38 % |
|  | MLP | 23 % |

In the functional evaluation test, we found that volunteers who attended our study were able to control the prothesis and execute the different hand gestures needed for ADL without any previous training. We also discovered that collecting data from participants' hands while they moved their hands and rotated their wrists (Figure 15) was an effective strategy to decrease the misclassification during changing the arm position, making this control system useful outside of laboratories. Moreover, the scores achieved by two volunteers in ARAT, showed that the developed SMG system to control the prosthetic hand has the potential to assist people with transradial hand amputation to perform different hand gestures needed for ADL (Figure 20), and the scores also proved that the functionality of the prosthesis is as good as a hand with moderate hand function. The B&B and TB&B tests showed the functionality of the developed robot with this novel controlling system in regard to manipulating objects using pinching. Moreover, in the experiments, no misclassification during hand movements, when volunteers wanted to transfer blocks, was observed.

During collecting data and testing the SMG controlling system, we noticed that gel pads and ultrasonic gels increase the possibility of probe movement, which significantly lowers the precision and reliability of the SMG controlling system. In addition to this, the skin will be in jeopardy due to the prolonged contact with moisture. Additionally, gels will contaminate the environment in which the ultrasound is mounted. Several potential solutions to these problems have been proposed and evaluated by researchers in the last few years. For instance, Wang et al. recently created a bioadhesive ultrasound (BAUS) device that can provide pictures from organs for 48 hours. To securely stick an array of piezoelectric elements to the skin without ultrasound gel, they utilized a soft, tough, anti-drying, and bioadhesive hydrogel-elastomer hybrid couplant layer [69]. In this study, we proposed to utilize a biocompatible sticky silicone pad as an alternative to ultrasound gel. It was discovered that a silicone pad has the potential to be used instead of ultrasound gel or gel pads, avoiding skin contact with moisture and thereby serious skin problems. We also observed that sticky silicone pads not only can help to capture images from muscle with a good resolution but also, by increasing the friction between the transducer and skin, prevent the relocation of the transducer, resulting in a decrease in misclassification in real-time control. In addition, during the offline test, the accuracy of classifying hand gestures in the able-bodied group was around 99% when ultrasound gel was used to collect the data. However, when ultrasound gel was replaced with a silicone pad the accuracy increased to 100%.

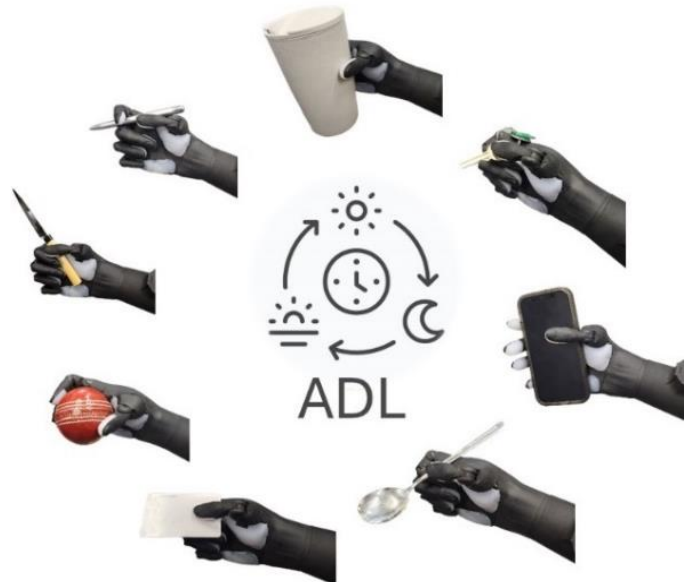

**Figure 20: ProRuka in activities of daily living:** Novel SMG system enabling multi-degrees of freedom prosthetic hand to be used in daily activities.

### 4.1. Limitations and future works

Even though the volunteers in this study were able to complete the various tasks, they found it difficult to pick up small objects due to a lack of sensory feedback. In the hand function test, they tried to control the prothesis only by looking at the hand movements without sensing the location of each finger, making it difficult for them to exercise excellent control over the prosthesis. Moreover, to develop a cost-effective prosthetic hand, a minimum of actuators and electronic items were used. However, in the hand function test, it was observed that it was difficult for participants to perform some daily activities due to the lack of wrist rotation. They could pick up blocks, but they needed to move their entire body to grasp and hold a glass, especially when simulating the pouring of water from one glass to another. Furthermore, sometimes participants complained about the prothesis blocking the view, making it difficult for them to see the objects they want to pick up. In addition, based on the tests results, it was observed that the prothesis could perform the pinch gesture, but it was difficult for the subjects to pick up small objects like coins, paper clips, ball bearings, etc. by pinching. Plus, to control the robot using the SMG technique, a wireless ultrasound transducer was mounted in the socket. The ultrasound used in this study weighed around 80 g, and to place it in the socket, we needed at least 110*56*10 mm3 of space, making the whole prosthesis bigger and heavier than other prosthetic hands with EMG sensors inside them.

In the future, different non-invasive methods for giving sensory feedback to amputees will be studied to not only increase the functionality of the hand but also decrease the phantom pain in people with hand amputations [70-74]. Moreover, by increasing the DOF of prosthesis and adding one more rotational joint in the thumb and one in the wrist, we will improve the dexterity, pinching, and wrist rotational movement of the prosthetic hand [75-77]. To remedy the limitations caused due to the rigidity of the prosthetic hand, in the future, a combination of rigid items and soft materials will be utilized to modify the prothesis and make it more like a human hand with higher dexterity and flexibility [72, 78, 79]. Finally, different AI methods will be used to predict not only the intended hand gestures, but also the amount of intended finger flexion. This will provide proportional and natural control over prosthetic hands.

### 4.2. Acknowledgements

We are really grateful to the subjects who participated in the research for their patience and commitment to our study. In addition, we would like to acknowledge and express our gratitude to May Wai Yoyo Lau for her help in facilitating training sessions, and Lyn Wong for her assistance in administrative support, as well as Yan To Ling, Tsung Kwan Shea, and Ka Shing Lee for their significant contributions they made to this project. Funding: This study was partially supported by Telefield Charitable Fund (ZH3V). Author contributions: Y-P.Z. and V.N. conducted the mechatronic integration, tested the device, and performed all the clinical trials. V.N. conducted the mechanical development of the systems. V.N. developed electronics. V.N. wrote the manuscript. Y-P.Z. and V.N. contributed to the writing of the paper. V.N. prepared the figures and videos. Y-P.Z. supervised the teams involved in the study and collected the funding to perform the study. All the authors have read and approved the manuscript in its final form. Competing **interests:** V.N., Y-P.Z., and T.K.S. are listed as inventors in the following patent application: US non-provisional patent, US serial no. 63/438,402, title: PROSTHETIC HAND DEVICE USING A WEARABLE ULTRASOUND MODULE AS A HUMAN MACHINE INTERFACE; submitted by the Spruson & Ferguson (Hong Kong) Limited, which cover the fundamental principles, controlling system and design of ProRuka. **Data and materials availability:** All data and materials are available in the main text.

### 5. REFERENCES

1. Damerla R, Qiu Y, Sun TM, Awtar S. A review of the performance of extrinsically powered prosthetic hands. IEEE Transactions on Medical Robotics and Bionics. 2021.
2. Cordella F, Ciancio AL, Sacchetti R, Davalli A, Cutti AG, Guglielmelli E, et al. Literature review on needs of upper limb prosthesis users. Frontiers in neuroscience. 2016;10:209.
3. Belter JT, Segil JL, SM B. Mechanical design and performance specifications of anthropomorphic prosthetic hands: a review. Journal of rehabilitation research and development. 2013;50(5):599.
4. Xu K, Guo W, Hua L, Sheng X, Zhu X, editors. A prosthetic arm based on EMG pattern recognition. 2016 IEEE International Conference on Robotics and Biomimetics (ROBIO); 2016: IEEE.
5. Laffranchi M, Boccardo N, Traverso S, Lombardi L, Canepa M, Lince A, et al. The Hannes hand prosthesis replicates the key biological properties of the human hand. Science robotics. 2020;5(46):eabb0467.
6. Biddiss EA, Chau TT. Upper limb prosthesis use and abandonment: a survey of the last 25 years. Prosthetics and orthotics international. 2007;31(3):236-57.
7. Resnik L, Meucci MR, Lieberman-Klinger S, Fantini C, Kelty DL, Disla R, et al. Advanced upper limb prosthetic devices: implications for upper limb prosthetic rehabilitation. Archives of physical medicine and rehabilitation. 2012;93(4):710-7.
8. Lewis S, Russold M, Dietl H, Kaniusas E. Satisfaction of prosthesis users with electrical hand prostheses and their sugggested improvements. Biomedical Engineering/Biomedizinische Technik. 2013;58(SI-1-Track-O):000010151520134385.
9. O'Neill C, editor An advanced, low cost prosthetic arm. SENSORS, 2014 IEEE; 2014: IEEE.
10. Gretsch KF, Lather HD, Peddada KV, Deeken CR, Wall LB, Goldfarb CA. Development of novel 3D-printed robotic prosthetic for transradial amputees. Prosthetics and orthotics international. 2016;40(3):400-3.

11. Kontoudis GP, Liarokapis MV, Zisimatos AG, Mavrogiannis CI, Kyriakopoulos KJ, editors. Open-source, anthropomorphic, underactuated robot hands with a selectively lockable differential mechanism: Towards affordable prostheses. 2015 IEEE/RSJ international conference on intelligent robots and systems (IROS); 2015: IEEE.
12. Ng KH, Nazari V, Alam M. Can Prosthetic Hands Mimic a Healthy Human Hand? Prosthesis. 2021;3(1):11-23.
13. Ke A, Huang J, Wang J, Xiong C, He J. Optimal design of dexterous prosthetic hand with five-joint thumb and fingertip tactile sensors based on novel precision grasp metric. Mechanism and Machine Theory. 2022;171:104759.
14. Mohammadi A, Lavranos J, Zhou H, Mutlu R, Alici G, Tan Y, et al. A practical 3D-printed soft robotic prosthetic hand with multi-articulating capabilities. PloS one. 2020;15(5):e0232766.
15. Yang D, Liu H. Human-machine shared control: New avenue to dexterous prosthetic hand manipulation. Science China Technological Sciences. 2021;64(4):767-73.
16. Furui A, Eto S, Nakagaki K, Shimada K, Nakamura G, Masuda A, et al. A myoelectric prosthetic hand with muscle synergy–based motion determination and impedance model–based biomimetic control. Science Robotics. 2019;4(31):eaaw6339.
17. Gu G, Zhang N, Xu H, Lin S, Yu Y, Chai G, et al. A soft neuroprosthetic hand providing simultaneous myoelectric control and tactile feedback. Nature Biomedical Engineering. 2021:1-10.
18. Weiner P, Starke J, Rader S, Hundhausen F, Asfour T. Designing prosthetic hands with embodied intelligence: The kit prosthetic hands. Frontiers in Neurorobotics. 2022;16:815716.
19. Nazarpour K. A more human prosthetic hand. Science Robotics. 2020;5(46):eabd9341.
20. Balasubramanian R, Santos VJ. The human hand as an inspiration for robot hand development: Springer; 2014.
21. Varol HA, Dalley SA, Wiste TE, Goldfarb M. Biomimicry and the design of multigrasp transradial prostheses. The Human Hand as an Inspiration for Robot Hand Development: Springer; 2014. p. 431-51.
22. Biagiotti L, Lotti F, Melchiorri C, Vassura G. How far is the human hand? a review on anthropomorphic robotic end-effectors. 2004.
23. Bicchi A, Gabiccini M, Santello M. Modelling natural and artificial hands with synergies. Philosophical Transactions of the Royal Society B: Biological Sciences. 2011;366(1581):3153-61.
24. Leo A, Handjaras G, Bianchi M, Marino H, Gabiccini M, Guidi A, et al. A synergy-based hand control is encoded in human motor cortical areas. Elife. 2016;5:e13420.
25. Weiner P, Starke J, Hundhausen F, Beil J, Asfour T, editors. The KIT prosthetic hand: design and control. 2018 IEEE/RSJ International Conference on Intelligent Robots and Systems (IROS); 2018: IEEE.
26. Murray CD. Embodiment and prosthetics. Psychoprosthetics: Springer; 2008. p. 119-29.
27. De Vignemont F. Embodiment, ownership and disownership. Consciousness and cognition. 2011;20(1):82-93.
28. D'Anna E, Valle G, Mazzoni A, Strauss I, Iberite F, Patton J, et al. A closed-loop hand prosthesis with simultaneous intraneural tactile and position feedback. Science Robotics. 2019;4(27):eaau8892.
29. Ritter H, Haschke R. Hands, dexterity, and the brain. Humanoid robotics and neuroscience: science, engineering and society. 2015.
30. Krebs HI, Palazzolo JJ, Dipietro L, Ferraro M, Krol J, Rannekleiv K, et al. Rehabilitation robotics: Performance-based progressive robot-assisted therapy. Autonomous robots. 2003;15(1):7-20.
31. Nazari V, Pouladian M, Zheng Y-P, Alam M. A compact and lightweight rehabilitative exoskeleton to restore grasping functions for people with hand paralysis. Sensors. 2021;21(20):6900.
32. Qassim HM, Wan Hasan W. A review on upper limb rehabilitation robots. Applied Sciences. 2020;10(19):6976.
33. Ribeiro J, Mota F, Cavalcante T, Nogueira I, Gondim V, Albuquerque V, et al. Analysis of man-machine interfaces in upper-limb prosthesis: A review. Robotics. 2019;8(1):16.
34. Begovic H, Zhou G-Q, Li T, Wang Y, Zheng Y-P. Detection of the electromechanical delay and its components during voluntary isometric contraction of the quadriceps femoris muscle. Frontiers in physiology. 2014;5:494.

35. Setiawan JD, Alwy F, Ariyanto M, Samudro L, Ismail R, editors. Flexion and Extension Motion for Prosthetic Hand Controlled by Single-Channel EEG. 2021 8th International Conference on Information Technology, Computer and Electrical Engineering (ICITACEE); 2021: IEEE.
36. Ahmadian P, Cagnoni S, Ascari L. How capable is non-invasive EEG data of predicting the next movement? A mini review. Frontiers in human neuroscience. 2013;7:124.
37. Gstoettner C, Festin C, Prahm C, Bergmeister KD, Salminger S, Sturma A, et al. Feasibility of a wireless implantable multi-electrode system for high-bandwidth prosthetic interfacing: animal and cadaver study. Clinical Orthopaedics and Related Research®. 2022;480(6):1191-204.
38. Taylor CR, Srinivasan SS, Yeon SH, O'Donnell M, Roberts T, Herr HM. Magnetomicrometry. Science Robotics. 2021;6(57):eabg0656.
39. Zheng Y-P, Chan M, Shi J, Chen X, Huang Q-H. Sonomyography: Monitoring morphological changes of forearm muscles in actions with the feasibility for the control of powered prosthesis. Medical engineering & physics. 2006;28(5):405-15.
40. Zhou Y, Zheng Y-P. Sonomyography: Dynamic and Functional Assessment of Muscle Using Ultrasound Imaging: Springer Nature; 2021.
41. Guo JY, Zheng YP, Huang QH, Chen X, He JF, Chan HL. Performances of one-dimensional sonomyography and surface electromyography in tracking guided patterns of wrist extension. Ultrasound Med Biol. 2009;35(6):894-902.
42. Chen X, Zheng YP, Guo JY, Shi J. Sonomyography (SMG) control for powered prosthetic hand: a study with normal subjects. Ultrasound Med Biol. 2010;36(7):1076-88.
43. Shi J, Chang Q, Zheng YP. Feasibility of controlling prosthetic hand using sonomyography signal in real time: preliminary study. J Rehabil Res Dev. 2010;47(2):87-98.
44. Shi J, Guo JY, Hu SX, Zheng YP. Recognition of finger flexion motion from ultrasound image: a feasibility study. Ultrasound Med Biol. 2012;38(10):1695-704.
45. Guo JY, Zheng YP, Xie HB, Koo TK. Towards the application of one-dimensional sonomyography for powered upper-limb prosthetic control using machine learning models. Prosthet Orthot Int. 2013;37(1):43-9.
46. Ma CZ-H, Ling YT, Shea QTK, Wang L-K, Wang X-Y, Zheng Y-P. Towards wearable comprehensive capture and analysis of skeletal muscle activity during human locomotion. Sensors. 2019;19(1):195.
47. Fifer MS, McMullen DP, Osborn LE, Thomas TM, Christie B, Nickl RW, et al. Intracortical Somatosensory Stimulation to Elicit Fingertip Sensations in an Individual With Spinal Cord Injury. Neurology. 2022;98(7):e679-e87.
48. Engdahl S, Mukherjee B, Akhlaghi N, Dhawan A, Bashatah A, Patwardhan S, et al., editors. A Novel Method for Achieving Dexterous, Proportional Prosthetic Control using Sonomyography. MEC20 Symposium; 2020.
49. Yang X, Yan J, Fang Y, Zhou D, Liu H. Simultaneous prediction of wrist/hand motion via wearable ultrasound sensing. IEEE Transactions on Neural Systems and Rehabilitation Engineering. 2020;28(4):970-7.
50. Akhlaghi N, Dhawan A, Khan AA, Mukherjee B, Diao G, Truong C, et al. Sparsity analysis of a sonomyographic muscle–computer interface. IEEE Transactions on Biomedical Engineering. 2019;67(3):688-96.
51. Li J, Zhu K, Pan L. Wrist and finger motion recognition via M-mode ultrasound signal: A feasibility study. Biomedical Signal Processing and Control. 2022;71:103112.
52. Guo J-Y, Zheng Y-P, Huang Q-H, Chen X. Dynamic monitoring of forearm muscles using one-dimensional sonomyography system. Journal of Rehabilitation Research & Development. 2008;45(1).
53. Castellini C, Passig G, Zarka E. Using ultrasound images of the forearm to predict finger positions. IEEE Transactions on Neural Systems and Rehabilitation Engineering. 2012;20(6):788-97.
54. Castellini C, Passig G, editors. Ultrasound image features of the wrist are linearly related to finger positions. 2011 IEEE/RSJ International Conference on Intelligent Robots and Systems; 2011: IEEE.
55. Nazari V, Zheng Y-P. Controlling Upper Limb Prostheses Using Sonomyography (SMG): A Review. 2023.
56. Nazari V, Zheng Y-P. Controlling Upper Limb Prostheses Using Sonomyography (SMG): A Review. Sensors. 2023;23(4):1885.

57. Engdahl SM, Acuña SA, King EL, Bashatah A, Sikdar S. First demonstration of functional task performance using a sonomyographic prosthesis: a case study. Frontiers in Bioengineering and Biotechnology. 2022;10:876836.
58. Xu Z, Todorov E, editors. Design of a highly biomimetic anthropomorphic robotic hand towards artificial limb regeneration. 2016 IEEE International Conference on Robotics and Automation (ICRA); 2016: IEEE.
59. Tilley AR. The measure of man and woman: human factors in design: John Wiley & Sons; 2001.
60. Mathiowetz V, Volland G, Kashman N, Weber K. Adult norms for the Box and Block Test of manual dexterity. The American journal of occupational therapy. 1985;39(6):386-91.
61. Kontson K, Marcus I, Myklebust B, Civillico E. Targeted box and blocks test: Normative data and comparison to standard tests. PloS one. 2017;12(5):e0177965.
62. Kontson K, Ruhde L, Trent L, Miguelez J, Baun K. Targeted Box and Blocks Test: Evidence of Convergent Validity in Upper Limb Prosthesis User Population. Archives of Physical Medicine and Rehabilitation. 2022;103(12):e132.
63. Administrasion USFaD. Targeted Box and Blocks Test (tBBT) 2022 [Available from: https://www.fda.gov/medical-devices/science-and-research-medical-devices/targeted-box-and-blocks-test-tbbt.
64. McDonnell M. Action research arm test. Aust J Physiother. 2008;54(3):220.
65. Yozbatiran N, Der-Yeghiaian L, Cramer SC. A standardized approach to performing the action research arm test. Neurorehabilitation and neural repair. 2008;22(1):78-90.
66. Buma FE, Raemaekers M, Kwakkel G, Ramsey NF. Brain function and upper limb outcome in stroke: a cross-sectional fMRI study. PLoS One. 2015;10(10):e0139746.
67. Ingram JN, Körding KP, Howard IS, Wolpert DM. The statistics of natural hand movements. Experimental brain research. 2008;188(2):223-36.
68. Kargov A, Pylatiuk C, Martin J, Schulz S, Döderlein L. A comparison of the grip force distribution in natural hands and in prosthetic hands. Disability and Rehabilitation. 2004;26(12):705-11.
69. Wang C, Chen X, Wang L, Makihata M, Liu H-C, Zhou T, et al. Bioadhesive ultrasound for long-term continuous imaging of diverse organs. Science. 2022;377(6605):517-23.
70. Middleton A, Ortiz-Catalan M. Neuromusculoskeletal arm prostheses: personal and social implications of living with an intimately integrated bionic arm. Frontiers in neurorobotics. 2020;14:39.
71. Raspopovic S, Valle G, Petrini FM. Sensory feedback for limb prostheses in amputees. Nature Materials. 2021;20(7):925-39.
72. Gu G, Zhang N, Xu H, Lin S, Yu Y, Chai G, et al. A soft neuroprosthetic hand providing simultaneous myoelectric control and tactile feedback. Nature biomedical engineering. 2023;7(4):589-98.
73. Marasco PD, Hebert JS, Sensinger JW, Beckler DT, Thumser ZC, Shehata AW, et al. Neurorobotic fusion of prosthetic touch, kinesthesia, and movement in bionic upper limbs promotes intrinsic brain behaviors. Science robotics. 2021;6(58):eabf3368.
74. Clemente F, D'Alonzo M, Controzzi M, Edin BB, Cipriani C. Non-invasive, temporally discrete feedback of object contact and release improves grasp control of closed-loop myoelectric transradial prostheses. IEEE Transactions on Neural Systems and Rehabilitation Engineering. 2015;24(12):1314-22.
75. Deijs M, Bongers R, Ringeling-van Leusen N, Van Der Sluis C. Flexible and static wrist units in upper limb prosthesis users: functionality scores, user satisfaction and compensatory movements. Journal of neuroengineering and rehabilitation. 2016;13:1-13.
76. Montagnani F, Controzzi M, Cipriani C. Is it finger or wrist dexterity that is missing in current hand prostheses? IEEE Transactions on Neural Systems and Rehabilitation Engineering. 2015;23(4):600-9.
77. Choi S, Cho W, Kim K. Restoring natural upper limb movement through a wrist prosthetic module for partial hand amputees. Journal of NeuroEngineering and Rehabilitation. 2023;20(1):135.
78. Zhao H, O'brien K, Li S, Shepherd RF. Optoelectronically innervated soft prosthetic hand via stretchable optical waveguides. Science robotics. 2016;1(1):eaai7529.

79.	Dunai L, Novak M, García Espert C. Human hand anatomy-based prosthetic hand. Sensors. 2020;21(1):137.